\documentclass[12pt]{article}

\parskip=\medskipamount            % space between paragraphs (LaTeX)
\def\7#1#2{\mathop{\null#2}\limits^{#1}}        % puts #1 atop #2

\def\beee{\begin{equation}}
\def\eeee{\end{equation}}
\def\dggg{^{\dagger}}
\def\bf{\mathbf{}}
\begin{document}
\bibliographystyle{unsrt}
 
\begin{center}
\textbf{Generalizations of Quantum Statistics}\\
[5mm]
O.W. Greenberg\footnote{email address, owgreen@umd.edu.}\\
{\it Center for Fundamental Physics\\
Department of Physics\\
University of Maryland\\
College Park, MD~~20742-4111, USA,\\
}
%University of Maryland Preprint PP-06-010}\\
~\\
\end{center}

\begin{abstract}

We review generalizations of quantum statistics,
including parabose, parafermi, and quon statistics, but
not including anyon statistics, which is special to 
two dimensions.

\end{abstract}                                       

The general principles of quantum theory allow statistics more general than
bosons or fermions. (Bose statistics and Fermi statistics are discussed
in separate articles.) The restriction to Bosons or Fermions requires
the symmetrization postulate,
``the states of a system containing $N$ identical particles
are necessarily either all symmetric or all antisymmetric under
permutations of the $N$ particles,'' or, equivalently, ``all states of identical 
particles are in one-dimensional representations
of the symmetric group~\cite{mes}.''  
A.M.L. Messiah and O.W. Greenberg discussed quantum mechanics 
without the symmetrization postulate~\cite{mesowg}.
The spin-statistics
connection, that integer spin particles are bosons and odd-half-integer spin
particles are fermions~\cite{pau}, is an independent statement. Identical
particles in 2 space dimensions are a special case, ``anyons.'' (Anyons are
discussed in a separate article.) Braid group statistics, a nonabelian analog
of anyons, are also special to 2 space dimensions 

All observables must be symmetric in the dynamical variables associated
with identical particles. Observables can not change the permutation symmetry
type of the wave function; i.e. 
there is a superselection rule separating states in inequivalent representations
of the symmetric group and when identical particles can occur in states 
that violate
the spin-statistics connection their transitions must occur in the same 
representation of the symmetric group. 
One can not introduce a small violation of
statistics by assuming the Hamiltonian is the sum of a 
statistics-conserving and a
small statistics-violating term, $H=H_S+\epsilon H_V$,
as one can for violations of parity, charge conjugation, etc.  
Violation of statistics must be introduced in a more subtle way.

S. Doplicher, R. Haag and J. Roberts~\cite{dop} classified
identical particle statistics in 3 or more space dimensions. 
They found parabose and parafermi statistics of positive integer orders, 
which had been introduced by H.S. Green~\cite{gre}, and
infinite statistics, which had been introduced by  
Greenberg~\cite{owg-q}. Parabose (parafermi) statistics allows up to $p$
identical particles in an antisymmetric state (symmetric) state.
Infinite statistics allows any number of identical particles in a symmetric or 
antisymmetric state.

Trilinear commutation relations,
\beee
[[a\dggg_k, a_l]_{\pm}, a\dggg_m]_-=2\delta_{lm}a\dggg_k
\eeee
with the vacuum condition, $a_k|0\rangle=0$, and 
single-particle condition, $a_k a\dggg_l|0\rangle=p \delta_{kl}|0\rangle$,
define the Fock representation of order $p$ parabose (parafermi) statistics.
Green found two infinite sets of solutions of these commutation rules, 
one set for each
positive integer $p$, by the ansatz,
\beee
a_k\dggg=\sum_{\alpha=1}^p b_k^{(\alpha) \dagger},~~a_k=\sum_{\alpha=1}^p b_k^{(\alpha)},
\eeee
where the $b_k^{(\alpha)}$ and $b_k^{(\beta) \dagger}$ 
are bose (fermi) operators
for $\alpha=\beta$ but anticommute (commute) for $\alpha \neq \beta$ for the 
parabose (parafermi) cases. The integer $p$ is the order of the parastatistics.  
For parabosons (parafermions) $p$ is the maximum number of
particles that can occupy an antisymmetric (symmetric) state.
The case $p=1$
corresponds to the usual Bose or Fermi statistics.
Greenberg and Messiah~\cite{owgmes} proved that Green's ansatz gives all 
Fock-like solutions of
Green's commutation rules. Local observables in parastatistics 
have a form analogous to the usual
ones; for example, the local current for a spin-1/2 theory is 
$j_{\mu}=(1/2)[\bar{\psi}(x), \psi(x)]_-$.  From Green's ansatz, it is clear
that the squares of all norms of states are positive; thus
parastatistics~\cite{del} gives a 
set of orthodox positive metric theories. 
Parabose or parafermi statistics for $p>1$ give gross 
violations
of Bose or Fermi statistics so that parastatistics theories are not useful to 
parametrize small violations of statistics.

The bilinear commutation relation 
\beee
a(k) a\dggg(l) - q a\dggg(l) a(k) =\delta(k,l),                          \label{q}
\eeee
with the vacuum condition, $a(k)|0\rangle =0$, 
define the Fock representation of quon statistics.
Positivity of norms requires $-1 \leq q \leq 1$~\cite{zag, spe}.  
Outside this range the squared norms become negative.
There is no commutation relation 
involving two $a$'s or two $a\dggg$'s.  There are $n!$ linearly independent $n$-particle
states in Hilbert space if all quantum numbers are distinct; these states differ only
by permutations of the order of the creation operators. 

For $q \approx \pm 1$, quons provide a formalism that can parametrize small 
violations of statistics so that quons are useful for quantitative tests
of statistics.
At $q= 1 (-1)$ only the symmetric (antisymmetric) representation of 
${\cal S}_n$ occurs. 
The quon operators interpolate smoothly between fermi and bose statistics 
in the
sense that as $q \rightarrow \mp 1$ the antisymmetric (symmetric) 
representations smoothly become more heavily weighted.  

Although there are $n!$ linearly independent
vectors in Fock space associated with a degree $n$ monomial in 
creation operators that
carry disjoint quantum numbers acting on the vacuum, there
are fewer than $n!$ observables associated with such vectors. 
The general observable
is a linear combination of projectors on the irreducibles of the symmetric
group. 

A convenient way to parametrize violations or bounds on violations of 
statistics uses the two-particle density matrix.  For fermions, 
$\rho_2=(1-v_F)\rho_a+v_F\rho_s$; for bosons, $\rho_2=(1-v_B)\rho_s+v_B\rho_a$.
In each case the violation parameter varies between zero if the statistics is
not violated and one if the statistics is completely violated.
R.C. Hilborn~\cite{hil2} pointed out that the transition matrix elements between
symmetric (antisymmetric) states are proportional to $(1 \pm q)$ so that the
transition probabilities are proportional to $(1 \pm q)^2$ rather than to 
$(1 \pm q)$.

Several properties of kinematically relativistic quon 
theories hold, including a generalization of Wick's theorem, 
cluster decomposition theorems and (at least for free quon fields)
the $CPT$ theorem;
however locality in the sense of the commutativity of observables 
at spacelike separation fails~\cite{owg-q}.  
The nonrelativistic form of locality
\beee
[\rho({\bf x}), \psi\dggg({\bf y})]_-=\delta({\bf x}-{\bf y}) \psi\dggg({\bf y}),
\eeee
where $\rho$ is the charge density, does hold.

Greenberg and Hilborn~\cite{owghil2} derived the generalization of the result due to 
E.P. Wigner~\cite{wig} and to P. Ehrenfest and J.R. Oppenheimer~\cite{ehr}
that a bound state of bosons and fermions is a boson unless it
has an odd number of fermions, in which case it is a fermion
generalizes for quons:  A bound
state of $n$ identical quons with parameter $q_{constituent}$ has parameter 
$q_{bound}=q_{constituent}^{n^2}$~\cite{owghil2}.

Note: References [1] through [14] are primary references. References [15] through [18]
are secondary references.

\end{document}